\documentclass[review,12pt]{elsarticle}

\usepackage{graphicx}

\usepackage{amssymb}

\journal{Nuclear Instruments and Methods in Physics Research Section A}

\begin{document}

\begin{frontmatter}



\title{A mobile antineutrino detector with plastic scintillators}


\author[minowa]{Y. Kuroda\fnref{JPSS}}
\ead{k@icepp.s.u-tokyo.ac.jp}

\author[minowa]{S. Oguri\fnref{JPSS}}
\author[minowa]{Y. Kato}
\author[minowa]{R. Nakata}
\author[icepp]{Y. Inoue}
\author[JAEA]{C. Ito}
\author[minowa]{M. Minowa}

\address[minowa]{Department of Physics, School of Science, University of Tokyo,
                 7-3-1, Hongo, Bunkyo-ku, Tokyo 133-0033, Japan}
\address[icepp]{International Center for Elementary Particle Physics,University of Tokyo,
                 7-3-1, Hongo, Bunkyo-ku, Tokyo 133-0033, Japan}
\address[JAEA]{Oarai Research and Development Center, Japan Atomic Energy Agency,
                 4002, Naritacho, Oarai, Ibaraki 311-1393, Japan}
\fntext[JPSS]{Research Fellow of the Japan Society for the Promotion of Science}

\begin{abstract}
We propose a new type segmented antineutrino detector made of plastic
 scintillators for the nuclear safeguard application.
A small prototype was built and tested to measure background events.
A satisfactory unmanned field operation of the detector system was
 demonstrated.
Besides, a detailed Monte Carlo simulation code was developed to estimate the
 antineutrino detection efficiency of the detector.
\end{abstract}


\end{frontmatter}


\section{Introduction}
\label{introduction}

Nuclear power plants are the most intense man-controlled sources of neutrinos. 
With an average energy of about 200\,MeV released per fission 
and about 6 neutrinos produced along the $\beta$-decay chain of the
fission products,
a total flux of $2\times10^{20}\nu/{\rm s}$ is emitted by a 1-GWt power plant.
Since unstable fission products are neutron-rich nuclei,
all $\beta$-decays are of $\beta^{-}$ type and
the neutrino flux is actually of pure electron antineutrinos ($\bar{\nu}_e$).
A prediction of its production rate and spectrum associated with the fission of ${}^{235}$U,${}^{238}$U,${}^{239}$Pu and ${}^{241}$Pu
exists and was recently revisited\cite{Vogel:1989}\cite{Mueller:2011}\cite{Huber:2011}.

The International Atomic Energy Agency uses an ensemble of procedures and technologies, collectively referred
to as the Safeguards Regime, to detect diversion of fissile materials from civil nuclear fuel cycle facilities into weapons programs.
One can ensure by measuring antineutrinos that a reactor was not stopped or started with unusual frequency.
Moreover, the rate and energy spectrum measurements 
enable a non-intrusive, real-time estimate of instantaneous and integrated thermal power 
and provide the information of the isotopic content of the in-core fuel.

Such a concept of antineutrino detection was first performed by a
Russian group at the Rovno Nuclear Power Plant in Ukraine\cite{Klimov:1994}.
Recently another experiment has been done at the San Onofre Nuclear Generating Station in the USA
using organic-liquid-scintillator detector with cubic meter scale
in an underground gallery\cite{bernstein:2002},\cite{bernstein:074905}.
They demonstrated a clear correlation between the changes in the reactor antineutrino
emission rate and the evolution of the reactor power and fissile inventory\cite{bowden:064902}.
These experiments have shown practical capabilities
suitable for near term use in the safeguards context.
In addition, a study of neutrino detection 
with a compact liquid scintillator detector deployed at the ground level
was recently carried out at the experimental fast reactor in Japan\cite{Furuta:2011}.

To improve the safety characteristics and make deployments and operations easier,
use of plastic detectors has some advantages.
Current liquid scintillators are usually based on 
organic solvents and therefore
have some security (toxicity and flammability) issues,
chemical compatibility with the container (they are good solvents) and need complicated retaining structures.

Additionally, detector mobility, simple installation and 
sea-level operation of detectors, rather than underground
have great impact for safeguards applications.

Taking above points into account, 
we propose a simple and mobile antineutrino detector,
``PANDA'', which stands for Plastic
Anti-Neutrino Detector Array.
The detector is made of plastic scintillators.
It can be installed on a dry van and 
placed outside of a reactor building (easily accessible zone).

Although these features have advantages in safeguards applications, there are some drawbacks.
Using plastic scintillators leads to some reduction in overall detection efficiency 
due to their smaller number of protons per volume 
comparing to typical liquid scintillators used in other neutrino experiments.
Furthermore, above-ground detection is a challenge because of the backgrounds induced by cosmic rays.
For inverse-beta detectors, the correlated background event rate at sea-level
may increase substantially relative to underground detectors.

In this paper, we report on the details of the
proposed PANDA detector and also the performance of a small prototype
called Lesser PANDA.

\section{Detector layout}
\label{detector}

\begin{figure}[t]
 \begin{center}
  \includegraphics[width=12cm]{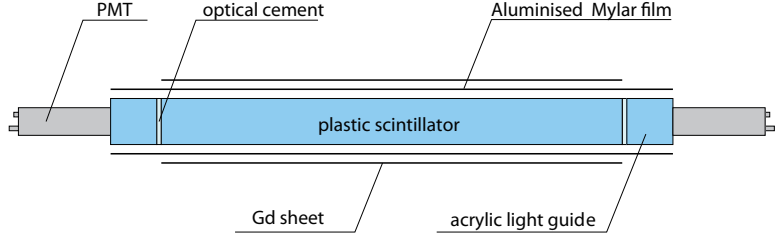}
  \caption{individual module structure}\label{fig:module}
 \end{center}
\end{figure}

The PANDA detector consists of modules made of 
plastic scintillator bars ($10 {\rm cm} \times 10 {\rm cm} \times 100 {\rm cm}$) wrapped with aluminized Mylar films
 and gadolinium (Gd) coated Mylar films (4.9 mg of Gd per ${\rm cm}^2$).
Each bar is connected to acrylic light guides and photomultipliers on 
both ends (Fig.~\ref{fig:module}).
The plastic scintillators are Rexon RP-408 
and the Gd coated films are manufactured by Ask Sanshin Engineering 
with their neutron shielding paint.
Hamamatsu H6410 (R329-02) PMTs are used for light 
detection.

Antineutrinos are detected via the inverse beta decay process: 
\begin{equation}\label{eq:inverse-beta}
 \bar{\nu}_e + p \rightarrow e^{+} + n
\end{equation}
and following reactions:
\begin{equation} \label{eq:positron_annihilation}
e^{+} + e^{-} \rightarrow 2 \gamma 
 {\rm ,}
\end{equation}
\begin{equation}\label{eq:neutron_capture_Gd155}
n + {}^{155}{\rm Gd} \rightarrow {}^{156}{\rm Gd}^* \rightarrow
 {}^{156}{\rm Gd} + \gamma\mbox{'s}
 {\rm ,}
\end{equation}
\begin{equation}\label{eq:neutron_capture_Gd157}
n + {}^{157}{\rm Gd} \rightarrow {}^{158}{\rm Gd}^* \rightarrow
 {}^{158}{\rm Gd} + \gamma\mbox{'s} 
 {\rm .}
\end{equation}
Here the antineutrino $\bar{\nu}_e$ 
interacts with the proton $p$ 
which is present in the plastic scintillator.
The neutron $n$ and the positron $e^{+}$ are detected by
delayed coincidence, 
providing a dual signature 
allowing strong rejection of the much more frequent singles backgrounds
due to natural radioactivity or cosmogenic neutrons.

The positron deposits energy via Bethe-Bloch ionization
as it slows in the scintillator, and upon annihilation with an electron,
it emits two gamma rays which can deposit 
additional energy of up to 1.022\,MeV (reaction (\ref{eq:positron_annihilation})). 
On the other hand the neutron carries away a few keV of energy from the antineutrino
interaction and is detected by capture in a Gd on the film
with an energy release of $\approx$ 8\,MeV via a gamma ray cascade
(reactions (\ref{eq:neutron_capture_Gd155}) and (\ref{eq:neutron_capture_Gd157})).
Reaction (\ref{eq:inverse-beta}) has an energy-dependent
cross section and a neutrino energy threshold of 1.806\,MeV.

\begin{figure}[t]
 \begin{center}
  \includegraphics[width=5cm]{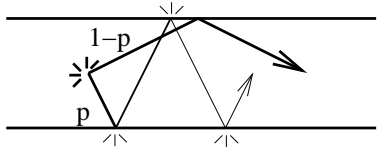}
  \caption{Light propagation model in the plastic scintillator bar}\label{fig:light_propagation}
 \end{center}
\end{figure}

\begin{figure}[t]
 \begin{center}
  \includegraphics[width=6cm]{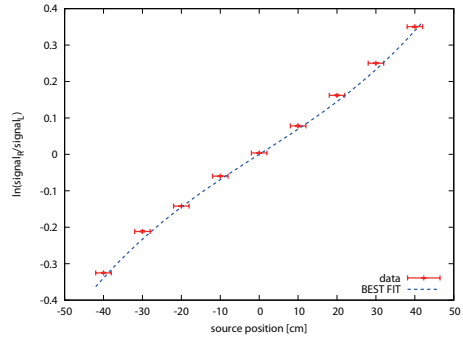}
  \caption{Logarithm of the ratio of PMT charge
  measured at one end to charge at the other end of the module
  vs. source position.
  The dashed curve is a two-parameter ($p$ and $l$ in 
  equation (\ref{eq:light_propagation})) fit of the ratio
  predicted by the light attenuation model.}
  \label{fig:light_proportion}
 \end{center}
\end{figure}

The light intensity ratio seen by each PMT pair allows one 
to estimate the position of the hit along the module.
Assuming the fraction $p$ of the light is attenuated in the plastic scintillator
exponentially as a function of distance from the point of emission to
the PMT and the other $(1-p)$ 
propagates practically without attenuation by
repeating total internal reflection at the surface
 (Fig.~\ref{fig:light_propagation}), 
the hit position and the light intensity are related through a simple equation:
\begin{equation}
 \label{eq:light_propagation}
 L_{\rm PMT} = L_{\rm emitted} ((1-p) + p\exp{(-x/l)})
  {\rm .}
\end{equation}
In this equation, $L_{\rm emitted}$ and $L_{\rm PMT}$ are the light
intensity emitted by the scintillator and reaching
the PMT,
$l$ is the attenuation length of scintillation light, and 
$x$ is the distance of the hit position from the PMT.
Fig.~\ref{fig:light_proportion} shows that the above relation (\ref{eq:light_propagation}) 
fits the data well.
Using the position of the hit and the charge outputs from each PMT,
one can estimate the energy deposit of the hit.
The position and energy resolutions which depend on the deposited energy
are about 30\,cm and 110\,keV for 500\,keV hit (10\,cm and 300\,keV for
4\,MeV hit) on the center respectively.

\begin{table}[t]
 \begin{center}
  \caption{Methods of Gd loading and neutron detection efficiency}\label{tab:neutronCapture}
  \begin{tabular}{p{12em}p{8em}p{8em}}
   \hline 
   methods & neutron captured by Gd[\%]& 
   mean time until the capture [$\mu$s]\\ \hline
   solution 0.1\%wt&     89.4& 28.4\\ \hline
   interleaving Gd foils and plastic scintillator sheets of: &&\\
   6\,cm t&          77.0& 54.0\\
   10\,cm t&         62.1& 94.9\\
   14\,cm t&         47.2& 138\\ \hline
   wrapping Gd foil over plastic scintillator bars of:&&\\
   6\,cm $\square$&   85.7& 29.6\\
   10\,cm $\square$&  76.0& 62.4\\
   14\,cm $\square$&  60.2& 95.6\\ \hline
  \end{tabular}
 \end{center}
\end{table}

The dependence of the Gd neutron capture efficiency 
on plastic scintillator granularity and shape were calculated 
using the simulation toolkit Geant4\cite{Agostinelli:2003},
which offers C++ libraries for simulating the passage of particles.
Three types of Gd loading methods were considered: 
dissolving Gd in plastic scintillators (0.1\%wt),
placing 25\,$\mu$m-thick Gd foils
($2.0\times10^{-2}{\rm g}/{\rm m}^2$)
between
scintillator plates (6\,cm, 10\,cm and 14\,cm thick) 
or wrapping square bars of plastic scintillators
(cross-sectional dimensions of $6 {\rm cm}\times 6 {\rm cm}$ , 
$10 {\rm cm}\times 10 {\rm cm}$ and $14 {\rm cm} \times 14{\rm cm}$) 
with 25\,$\mu$m-thick Gd foils.
In the simulation, the total dimensions in three axes were taken to be
infinite.

Ten kiloelectronvolt neutrons
were generated at a single point inside of the plastic
 scintillator, and the probability of capture by Gd and the mean time until the
 capture were estimated.
The results of this study are presented in Table
\ref{tab:neutronCapture}.
It should be noted here that all neutrons were
captured by nuclei of hydrogen or gadolinium.

The simulation result showed that 
the segmented scintillators interleaved with the gadolinium foils have 
performance comparable to the gadolinium loaded plastic scintillators.
Moreover the technique to produce clear and colorless Gd loaded plastic scintillators 
is less established. 
Therefore we did not choose to load gadolinium in scintillators.

Eventually we selected the square bar type module which has an advantage 
in detecting the positions of energy deposit 
and therefore energy spectrum of the positrons,
which is related directly to the neutrino energy spectrum.
The granularity of the modules 
was determined to be
$10\,{\rm cm}\times 10\,{\rm cm}$ 
taking costs into consideration.

\begin{figure}[t]
 \begin{center}
  \includegraphics[width=8cm]{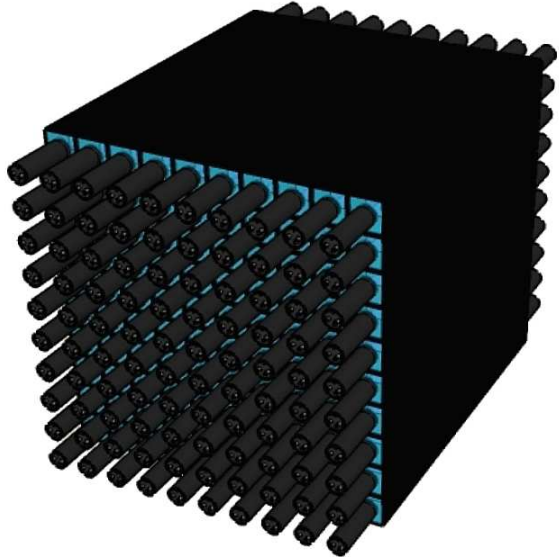}
  \caption{The concept design of the PANDA antineutrino detector.
  The approximate total target size is $1{\rm\, m}^3$.}
  \label{fig:PANDA_concept}
 \end{center}
\end{figure}

The design of the PANDA detector is shown in Fig.~\ref{fig:PANDA_concept}.
The detector consists of one hundred identical modules 
described above in this section.
As it has fine segmented structure,
it becomes possible to use the event topology information 
to tag anti-neutrino events and to discriminate them from background.

The fine segmentation of the detector also gives the capability to 
identify and reject passing tracks such as cosmic ray muons.
Passing muons are characterized by series of hits distributed
along a line in the detector in addition to its relatively large energy deposit.
The characteristics helps the PANDA detector
to be designed without additional veto counters surrounding the whole detector.

\begin{figure}[t]
 \begin{center}
  \includegraphics[width=12cm]{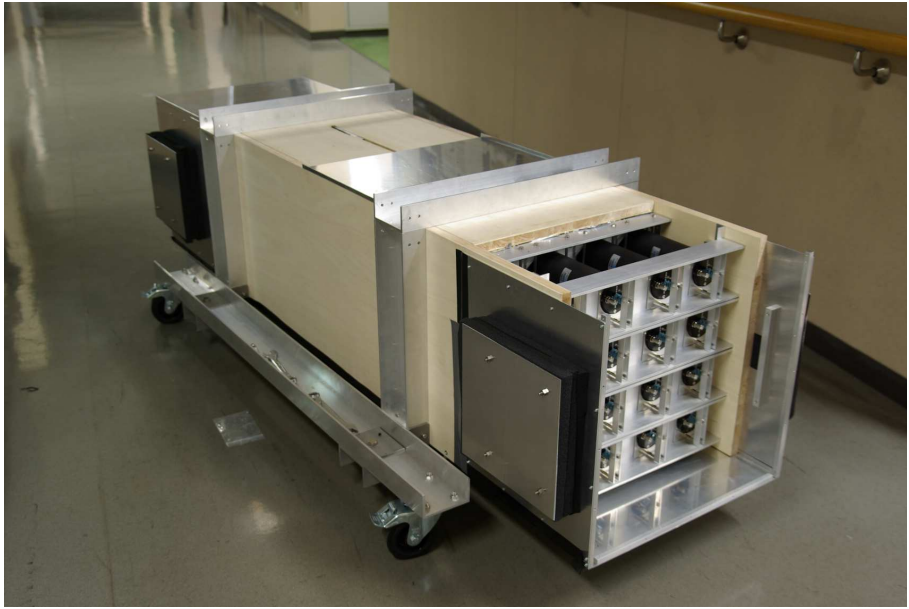}
  \caption{The prototype detector, Lesser PANDA,
  consists of 16 modules.}\label{fig:LesserPANDA}
 \end{center}
\end{figure}

As a first step before building the full-size PANDA detector, a smaller prototype
``Lesser PANDA'' with 16 modules (Fig.~\ref{fig:LesserPANDA}) 
was constructed and was tested by 
measuring environmental radiation at a commercial reactor.
This prototype was used to verify the background rejection ability of its
modular configuration with above-ground condition.

\section{Electronics and Data Acquisition System}
\label{daq}

\begin{figure}[t]
 \begin{center}
  \includegraphics[width=10cm]{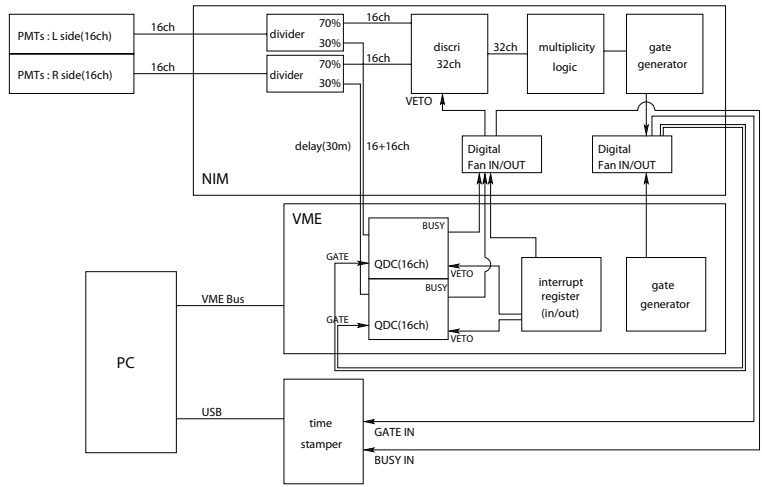}
  \caption{The data acquisition system and trigger logic}\label{fig:daq}
 \end{center}
\end{figure}

\begin{figure}[t]
 \begin{center}
  \includegraphics[width=5cm]{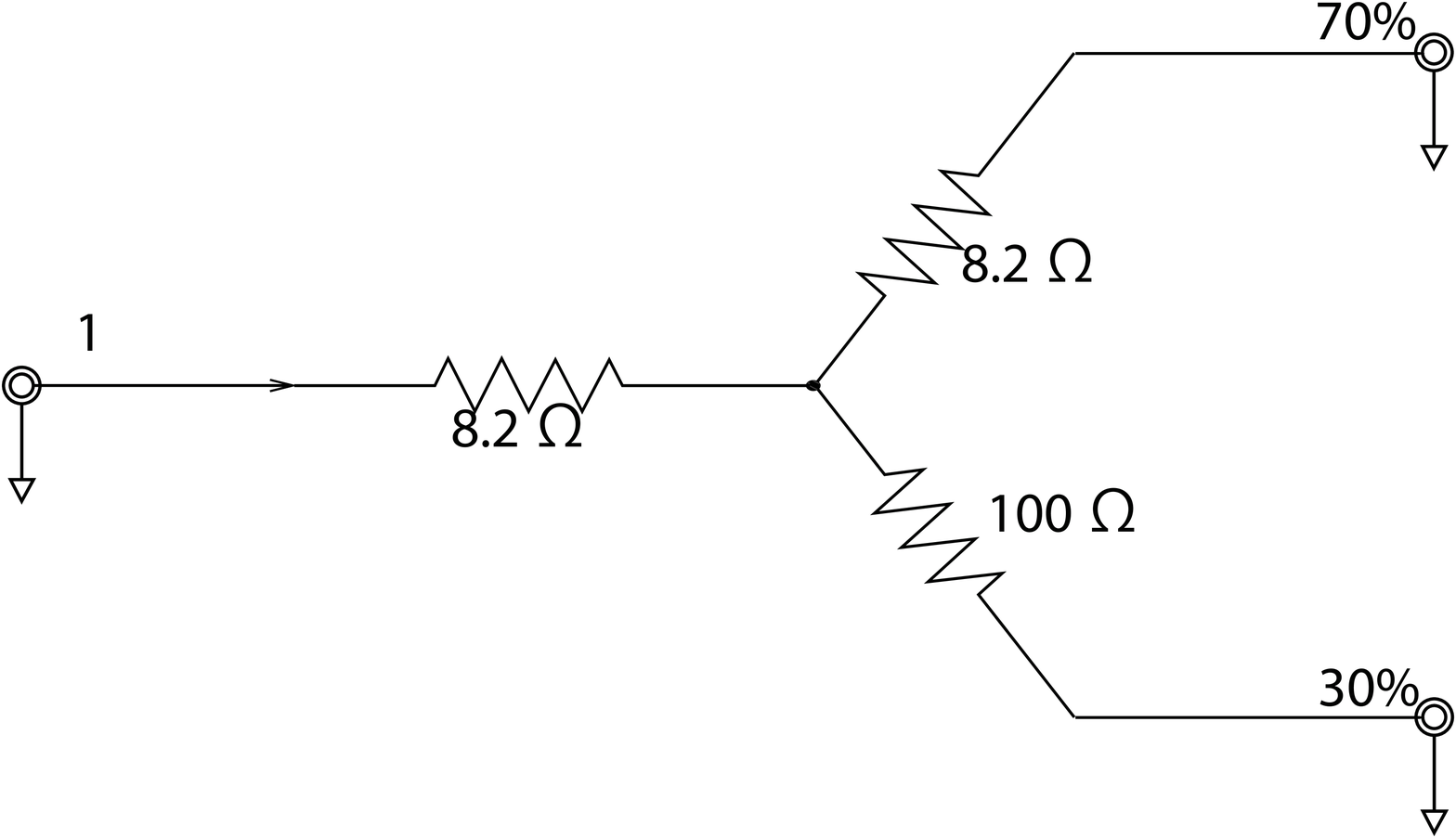}
  \caption{Circuit schematic of signal divider}\label{fig:divider}
 \end{center}
\end{figure}

Fig.~\ref{fig:daq} shows the 
block diagram
of the data acquisition and trigger system
for Lesser PANDA. 
Each PMT signal 
was 
divided into two (Fig.~\ref{fig:divider}): 
30\% of the original 
charge was 
sent to CAEN V792 multievent
Charge-to-Digital-Converters (QDCs)
and the other 70\% was 
passed to Technoland Corporation N-TM 405 leading edge discriminators.
Hardware thresholds were set at the lower limit of N-TM 405 which were
equivalent to about 150\,keV 
of energy deposit in
the plastic scintillators 
at the far end of the corresponding
PMTs.

Gates for the QDCs were produced by the following trigger logic.
A hit multiplicity was counted by a REPIC RPN-130 multiplicity logic
module and
 a logic pulse was sent to a gate generator whenever 
four or more PMTs fired in coincidence with one another
giving typical trigger rate of 2.7\,kHz.
Non-retriggerable logic pulses of 400\,ns duration were produced here.
The coincidence width of the multiplicity logic was 50\,ns.
Although we did not require the triggers to be both pairs of PMTs
on two or more scintillator modules, the two-pair condition was
applied in the analysis phase.
Corresponding to each trigger, the charge outputs of all PMTs were
recorded by the QDCs.
Meanwhile the trigger timing was recorded by a custom-made time stamper.

We set up periodic pauses of the data acquisition 
to assure the synchronism of the data from all the QDCs and the time stamper 
was retained.
During the pauses, 
the data acquisition system counted the number of the collected data 
and when any mismatches between the QDCs or the time stamper were found, 
the system discarded all the data after the previous synchronization process.

Due to the conversion times of the QDCs (about $5.7\mu$s) and synchronization
process, the DAQ system had about 3.5$\%$ dead time when the trigger rate
was 2.7 kHz.
The QDCs have 32 event buffer memories and they incur little dead time as long
as the read out system can sustain the data rate.
The amount of data transferred from QDCs to PC was 
small enough ($\approx 370$ kByte/sec) at 
the assumed trigger rate;
therefore the read out dead time was negligible ($<0.01\%$).

The time stamper was
implemented on a Xilinx
Spartan-6 FPGA running at 50-MHz clock.
The trigger timing information allows to calculate the 
interval
between a pair of events.
The time stamper also recorded the timing of the busy signals from QDCs
which was used to calculate the dead time of the
measurements.
These timing information was stored in a cache memory on the FPGA once 
and sent to a hard disk drive sequentially via USB~2.0 (Universal Serial Bus 2.0).
The data transfer speed of USB~2.0 is high enough
to run without dead time due to the time stamper at trigger rate of 2.7kHz.

\begin{figure}[t]
 \begin{center}
  \includegraphics[width=10cm]{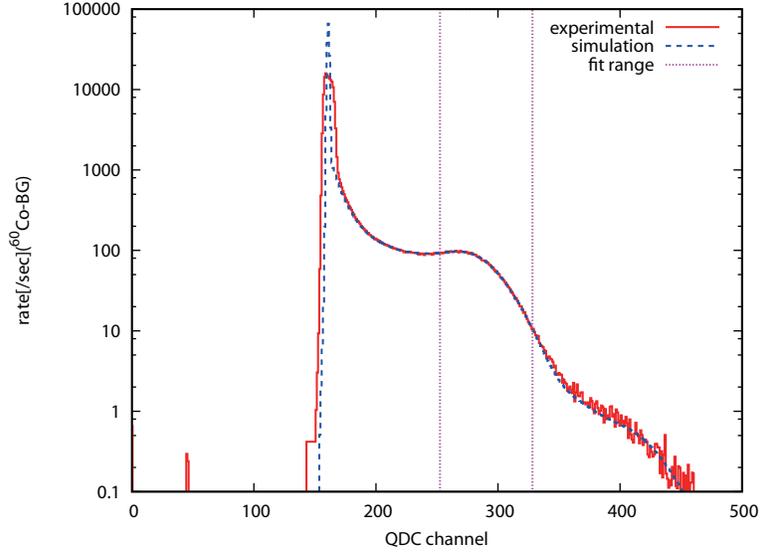}
  \caption{An example of 
  energy spectrum in one module obtained with ${}^{60}$Co calibration
  source after background subtraction,
  compared with the Geant4 simulation (dashed
  line).}\label{fig:calibration}
 \end{center}
\end{figure}

Energy calibration of the detector was carried out using
a ${}^{60}$Co gamma-ray source temporarily placed at
three different positions: the very center of the detector or the
centers of both the PMT sides.
Calibration constants were evaluated for each module separately.

Multiple Compton scattering of the gamma-rays with energies of 1.17\,MeV and
1.33\,MeV in the detector
or in the surrounding materials and
statistical and electronic fluctuation of the output signals 
produce a broad spectral feature in the measured energy spectrum
(Fig.~\ref{fig:calibration}).
To compare with the measured energy spectrum,
we performed a Monte Carlo simulation
of gamma-ray transports in the detector using Geant4\cite{Agostinelli:2003}.

The calibration procedure was as follows:
we took data with a ${}^{60}$Co source;
determined the values of the pedestals;
fitted the broad spectral feature by Compton scattering of
${}^{60}{\rm Co}$ gamma-rays
of the simulation to the three data
at the same time taking into account
the light attenuation described in Section~\ref{detector}.

When the detector was in operation, relative gain drift of each module was corrected using gamma-rays 
arising from natural radioactivity.
The background gamma-ray spectrum was taken just after the first calibration
as a reference and each data set was compared to it.
The gain drift of each PMT was 5.1\% on root mean square and corrected every about 6 minutes.

Also the pedestals were monitored.
The values of the pedestals were 
determined from each data set. 
Since the signals from all modules were digitized simultaneously
when any four or more PMTs fired, 
a large number of recorded events for particular modules had zero
energy
and were a measurement of the energy pedestal.

The DAQ systems were installed on a dry van with the detector.
NTT DoCoMo cell phone network was used 
for remote monitoring of the detector health status.
The remote connection to the system was also used 
to control the detector system from a distance.
The Lesser PANDA system typically operated 
without physical intervention for two weeks.

\section{Detector Simulation}
\label{result_simulation}

We used Monte Carlo simulations to have insight into the 
particle transport properties of the PANDA/Lesser PANDA detectors and to
estimate the $\bar{\nu}_e$ detection efficiencies of our detectors
taking advantage of the fine segmentation property to reject the
background events.

\begin{figure}[t]
 \begin{center}
  \begin{minipage}{0.45\hsize}
   \begin{center}
    \includegraphics[width=6cm]{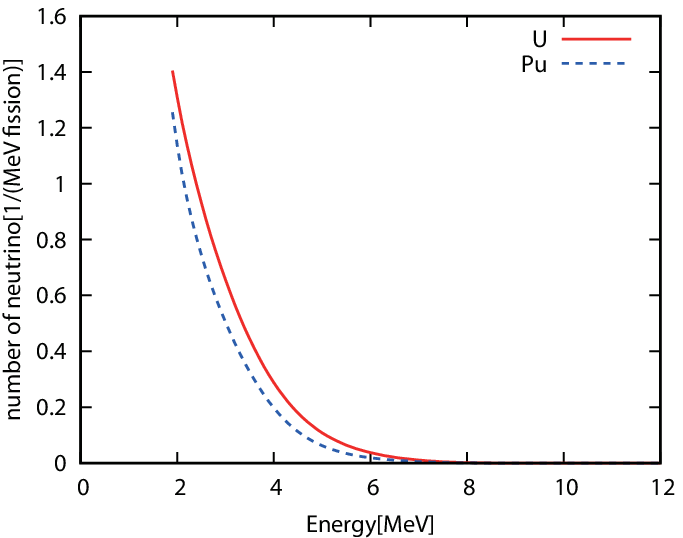}
    (a)
   \end{center}
  \end{minipage}
  \hspace{0.02\hsize}
  \begin{minipage}{0.45\hsize}
   \begin{center}
    \includegraphics[width=5.5cm]{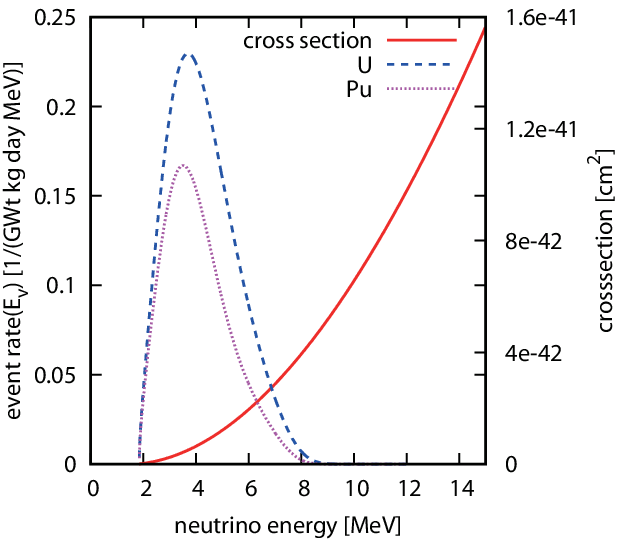}
    (b)
   \end{center}
  \end{minipage}
  \caption{(a): The antineutrino spectra from
  ${}^{239}$Pu and ${}^{235}$U as emitted by a reactor.
  (b): Inverse beta decay 
  cross section and detected antineutrino spectra 
  which is the product of the emitted spectra and the cross section.
  It assumes that a 1-kg plastic scintillator detector at a standoff of
  30\,m from the reactor core.}\label{fig:ibd_eventrate}
 \end{center}
\end{figure}

The plot on (a) in Fig.~\ref{fig:ibd_eventrate} shows the
predicted energy spectra of antineutrinos per fission for
the two most important fissile isotopes ${}^{239}$Pu and
${}^{235}$U\cite{Vogel:1989}
The plot (b) shows the same spectra after folded with the
energy-dependent cross section of the inverse beta decay.
The folding is done by the equation:
\begin{equation}
 N_{\bar{\nu},i}(E_{\bar{\nu}}) = \frac{TN_p}{4\pi D^2}\sigma(E_{\bar{\nu}})\phi_i(E_{\bar{\nu}}).
\end{equation}
In this equation, $N_{\bar{\nu},i}(E_{\bar{\nu}})$ is the number of
antineutrino interactions per MeV in the detector in a time interval $T$ 
from the $i$-th isotope. In practice the index runs over the four dominant
fissioning isotopes ${}^{235}$U, ${}^{238}$U, ${}^{239}$Pu and
${}^{241}$Pu, which account for $\sim$99.9\% of fissions in a typical reactor.
$N_p$ is the number of target protons in the detector, $D$ is the
distance from the detector to the center of the reactor core.
$\sigma$ and $\phi_i$ are the cross section for the inverse beta
interaction and the antineutrino number density per MeV and second for the $i$-th
isotope respectively.
A cubic-meter-sized plastic scintillator detector at a standoff of
$30\,{\rm m}$ from a reactor core is expected to measure 
$\sim 2.3 \times 10^3$ events per day, 
for $3$ GWt power plant
and a perfectly efficient detector.

We simulated positrons and neutrons produced via the inverse beta decay
events uniformly distributed  throughout the plastic scintillator volume, 
and recorded the total visible energy in each module.
The energy and angular distributions of positrons and neutrons were derived
from the inverse beta decay angular-differential cross section depending on
the antineutrino energy $E_{\bar{\nu}}$ \cite{Vogel:1999} and spectra shown in
Fig.~\ref{fig:ibd_eventrate}.

We used the Geant4 simulation toolkit\cite{Agostinelli:2003}
for simulating the passage of particles.
To account for the optical photon response, we utilized the simple light
propagation model described in Section~\ref{detector}.
We simulated particle transport in the scintillators,
the surrounding materials including the acrylic light guides, 
the PMTs, the aluminized Mylar films
and the gadolinium coated Mylar films.
Especially for the Lesser PANDA detector,
the transport in the outer aluminum and wooden support structures were also
simulated.

We modeled the resolution of the digitized signal $\sigma_{\rm sig}(E)$ as
\begin{equation}
 \sigma_{\rm sig}(E) = 
  \sqrt{ \sigma^2_{\rm electronics} + \sigma^2_{\rm p.e.}(E)}.
\end{equation}
Here, the constant $\sigma_{\rm electronics}$ represents the effect of the electric
noise of the data acquisition system.
It was determined by fitting the width of the pedestal distribution.
$\sigma_{\rm p.e}(E) \equiv k\sqrt{E}$ is the variation in response
caused by the finite number of photoelectrons generated in the event.
Coefficient $k$ for each module 
came out of the module calibration described in
Section~\ref{daq}.

\begin{table}
 \begin{center}
  \caption{The Antineutrino Event Selection Criterion}\label{tab:selection_cuts}
  \begin{tabular}{cc}
   \hline
   Prompt Energy& $ 3\,{\rm MeV} \leq E_{\rm total} \leq 10\,{\rm MeV}$ \\
                & $ E_{\rm 1st} \leq 6\,{\rm MeV}$  \\ 
                & $ 200\,{\rm keV} \leq E_{\rm 2nd} \leq 500\,{\rm keV}$  \\ \hline
   Delayed Energy& $ 3\,{\rm MeV} \leq E_{\rm total} \leq 8\,{\rm MeV}$ \\
                & $ E_{\rm 1st} \leq 5\,{\rm MeV}$  \\ 
                & $ 500\,{\rm keV} < E_{\rm 2nd}$  \\ \hline
   Prompt-delayed interval& $ 6\mu{\rm s} < t_{\rm int} < 200\mu{\rm s} $ \\
   \hline
  \end{tabular}
 \end{center}
\end{table}

\begin{table}
 \begin{center}
  \caption{Detection Efficiency}\label{tab:detection_efficiency}
  \begin{tabular}{ccc}
   \hline
                        &Lesser PANDA      &PANDA\\ \hline
   Target mass          &160 kg           & 1000 kg  \\
   Detection efficiency &4.0\%            & 11.6\%    \\
   \shortstack{expected neutrino event rate \\ (at 30m from a 3 GWt reactor)}
                        &15/day    & $2.7 \times 10^2$/day\\
   \hline
  \end{tabular}
 \end{center}
\end{table}

\begin{table}
 \begin{center}
  \caption{Detection Efficiency of Each Stage of the Criteria}\label{tab:part_efficiency}
  \begin{tabular}{ccc}
   \hline
   Quantity & Efficiency \\
   \hline
   Prompt             & 12\% \\
   Delayed            & 23\% \\
   time interval cut  & 89\% \\
   \hline
   Total    & 4.0\% \\
   \hline
  \end{tabular}
 \end{center}
\end{table}

We applied the selection cuts shown in Table \ref{tab:selection_cuts} to
the simulation outputs and estimated the detection efficiency.
Prompt and delayed events were selected using 
the sum of the energy deposits of all the modules ($E_{\rm total}$),
the highest and the second highest
values among all the modules ($E_{\rm 1st},E_{\rm 2nd}$).

The threshold for the prompt total energy $E_{\rm total}$ was placed at 3\,MeV 
so that all background gamma-rays were rejected here.
The energy distribution of the antineutrino-generated positrons is directly
related to the antineutrino energy distribution 
therefore no events with $E_{\rm total}$ beyond 10\,MeV (6\,MeV for $E_{\rm 1st}$) 
were admitted 
as extremely few antineutrino interactions were expected above these
energies.
The $E_{\rm 2nd}$ condition was set to tag the gamma rays created via
annihilation of positrons.

The delayed total energy $E_{\rm total}$ threshold was
also placed at 3\,MeV. 
This threshold was set 
as high as possible to reject background gamma rays
and low enough to retain as many delayed neutron events 
with cascading gamma rays from neutron capture by Gd
as possible.
Events with delayed energy of greater than 8\,MeV were also excluded, as
the predicted delayed energy was smaller than it.
The $E_{\rm 2nd}$ threshold for delayed events was placed at 500 keV.
Since more than one gamma rays with few MeV of energies are emitted from
neutron capture by Gd, there are comparatively large number of events 
with $E_{\rm 2nd}$ greater than 500 keV.
For the same reason, 
it is unlikely for the gamma rays to deposit 
such a large quantity of energy in one module.
Therefore events with $E_{\rm 1st}$ greater than 5\,MeV were discarded.

All values for the energy criteria were finally verified comparing 
energy spectra from the detector simulation and the background measurement.
The detection efficiencies were comparatively insensitive to the energy cuts
around these thresholds. 

The detection efficiency is also affected by the prompt-delayed interval cut.
The interval distribution of antineutrino interactions is 
described by an exponential decay.
According to the simulation result, the time constant of the exponential is 
about 60\,$\mu$s and 
the interval cut of 200\,$\mu$s means that
only about 10\,\% of the neutrino events are lost by this selection.

Table~\ref{tab:detection_efficiency} shows the resulting total detection
efficiencies and the expected event rates for the Lesser PANDA and the PANDA detectors.
Table~\ref{tab:part_efficiency} shows the estimated efficiencies of each 
stage of our selection criteria.

\begin{figure}[t]
 \begin{center}
  \begin{minipage}{0.48\hsize}
   \begin{center}
    \includegraphics[width=7cm]{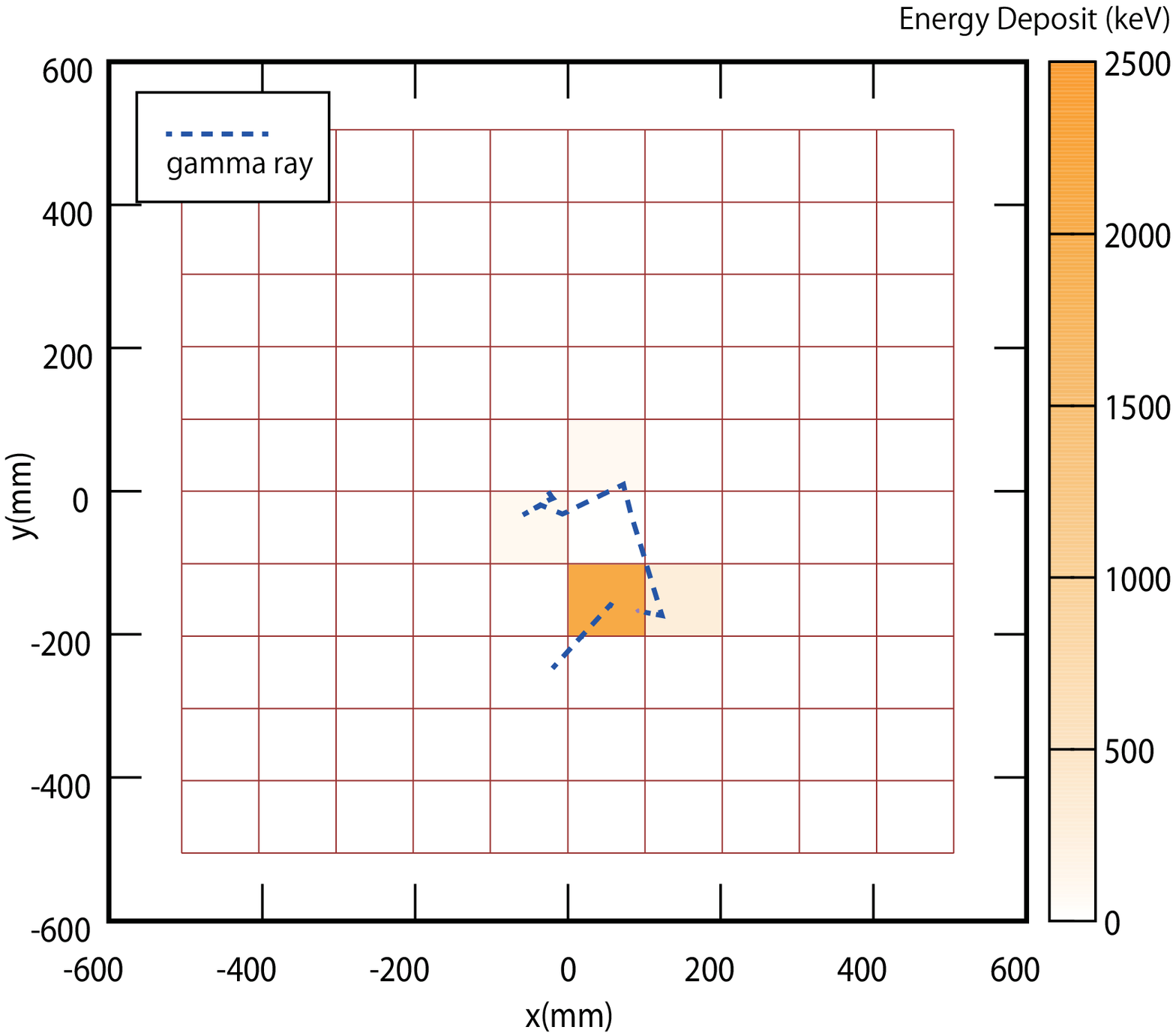}
    (a)
   \end{center}
  \end{minipage}
  \hspace{0.02\hsize}
  \begin{minipage}{0.48\hsize}
   \begin{center}
    \includegraphics[width=7cm]{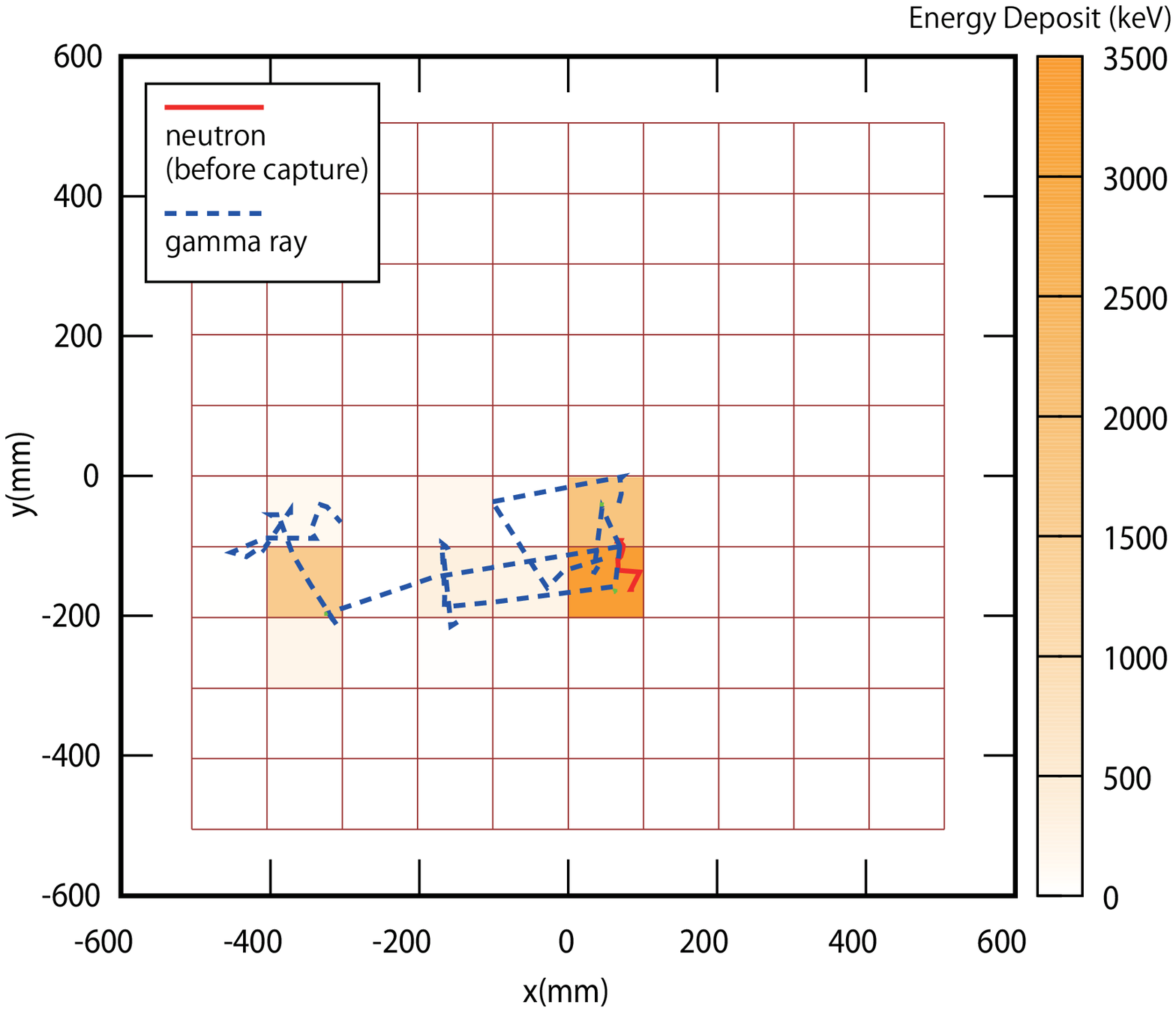}
    (b)
   \end{center}
  \end{minipage}
  \caption{(a): Typical particle trajectories 
  and energy deposit
  in the full size PANDA
  detector of a simulated antineutrino prompt event seen from the
  PMT side of the detector. 
  (b): Corresponding delayed event.}\label{fig:simulated_event}
 \end{center}
\end{figure}

Fig.~\ref{fig:simulated_event} shows a typical simulated antineutrino
event detected by the full size PANDA detector.
A hit by a positron and one of two annihilation gamma rays are seen in
the prompt event.
Multiple hits of gamma rays emitted by Gd are also seen in the delayed
event after 13 $\mu$s of the prompt event.

\section{Background measurement}
\label{result_experiment}

Ahead of the measurement at the commercial reactor, 
the Lesser PANDA detector was
tested for two days in the van near the university building for
background measurement. 
We counted the number of neutrino-like background events which are
hardly distinguishable from the neutrino interaction and 
it amounted $981 \pm 82$\,events/day.

The number of neutrino-like events were calculated as follows.
First, we counted the number of events 
which satisfied all the selection criteria described in
Table~\ref{tab:selection_cuts}.
Then we counted the number of accidental coincidences by 
applying the same criteria except for the prompt-delayed interval, 
which was changed to $1006\mu{\rm s} < t_{\rm int} < 1200\mu{\rm s}$.
The difference of these two numbers represents 
the number of events whose prompt and delayed signals had 
some correlation.
We considered these correlated events as the neutrino-like backgrounds.
The operation did not reduce the detection efficiency significantly.
The subtraction of the accidental coincidence was important for our analysis
because of fluctuating gamma-ray background 
especially during the measurement at the Hamaoka Nuclear Power Plant described below
due to radioactive dust which
originated in the Fukushima-1 nuclear power plant accident.

The detector was then
deployed at Unit 3 of the Hamaoka Nuclear Power Plant
of Chubu Electric Power Co., Inc.
There are three operational reactors in this station; all of them are
boiling water reactors and have maximum thermal (electric) power of
3.3--3.9GWt (1.1--1.4GWe) each.
The detector was located by the reactor building of Unit 3 (3.3GWt) 
at a distance of 39.8\,m from the reactor core. 
The Unit 3 reactor was under scheduled inspection and was not in operation then
\footnote{
We intended to measure the event rate change before and after a start up
of the reactor when the inspection were over, but it did not start because
of the general shutdown request by the Prime Minister after the 
Fukushima-1 nuclear power plant accident.
}.

During a 70-day period of the background measurement,
the average daily rate of neutrino-like events
was distributed at $888\pm 70$\,events/day. 
The standard deviation is somewhat greater than the statistics due to as
yet unknown reason.
These events are comprised of non-antineutrino backgrounds.
We observed no significant excess in the count rate
compared to the rate measured near the university building.

\begin{figure}[t]
 \begin{center}
  \includegraphics[width=8cm]{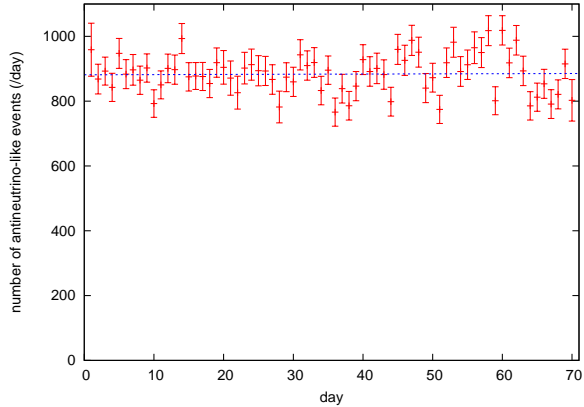}
  \caption{Daily number of correlated background 
  events which satisfied the
  antineutrino selection criteria.
  Error bars are only statistical.}\label{fig:event_per_day}
 \end{center}
\end{figure}

A linear fit to 
the daily detection rate of antineutrino like events 
during reactor off period plotted versus day has a slope of 
$0.06 \pm 0.36 \, {\rm events}/{ \rm day}^2$
 (Fig.~\ref{fig:event_per_day}), which is consistent with a constant background.

The van with the detector on it was parking as close as possible to the
reactor building.
There were no connections between the detector and the reactor building
except a 100-volt AC power line,
which supplied about 600 W of electricity for the whole detector system.

A hard disk drive with event data was replaced every two weeks and was
carried back to Tokyo for the analysis.
The detector system was otherwise operated without 
a human intervention throughout the period, 
and its status was monitored from the university in Tokyo via
the cell phone network.

\section{Discussion and conclusions}
\label{discussion}

We proposed a new type of segmented antineutrino detector made of one
hundred plastic scintillator bars wrapped in Gd-coated films for the
nuclear safeguard application.
The detector is all solid states and nonflammable.
It is comparatively small, can be carried easily in a van 
and is intended to monitor a reactor from outside of the reactor building.

As a first step, a smaller prototype was built and tested near the
university building and also near the reactor building of the nuclear
power plant.
The measured background rates were comparable on both the sites.
We have shown that an unmanned operation is possible for a few months
only with biweekly hard disk drive exchanges for the data collection.

A detailed Monte Carlo simulation code was developed for both the full
size and prototype detectors. A detection efficiency of
11.6\% and 4.0\% were estimated for the full size and prototype
detectors, respectively.
Clearly, it would be hard to detect statistically significant number of
antineutrino events with the prototype detector because of its low
detection efficiency and relatively high measured background rate
although current selection criteria are not fully optimized yet.

With the full-size detector of effective target mass of 1 ton, we have a good
chance to detect sufficient number of antineutrino events at reasonable
standoff from commercial nuclear power stations.

\section{Acknowledgements}
\label{acknowledgements}

The authors thank Chubu Electric Power Co.,Inc for its cooperation for
our experiment on site of Hamaoka Nuclear Power Plant.

This research was partially supported by the 
Japanese Ministry of Education, Science, Sports and Culture, 
Grant-in-Aid for COE Research,
Grant-in-Aid for Scientific Research (B), 
and Grant-in-Aid for JSPS Fellows
and also by the Mitsubishi Foundation.





\bibliographystyle{model1a-num-names}
\bibliography{k_bib}







\end{document}